\documentclass[pra,aps,showpacs]{revtex4}
\usepackage{amsmath,amssymb}
\usepackage{graphicx}

\newcommand{\non}{\nonumber}
\newcommand{\myscalebox}[1]{\scalebox{0.6}[0.6]{#1}}

\begin{document}

\title{Towards Finite-Dimensional Gelation}

\author{Kurt Broderix\footnote{deceased}, Martin Weigt, and Annette
  Zippelius} 

\affiliation{Institut f\"ur Theoretische Physik, Universit\"at
  G\"ottingen, Bunsenstr. 9, D-37073 G\"ottingen, Germany}

\date{\today}

\begin{abstract}
  We consider the gelation of particles which are permanently
  connected by random crosslinks, drawn from an ensemble of
  finite-dimensional continuum percolation. To average over the
  randomness, we apply the replica trick, and interpret the replicated
  and crosslink-averaged model as an effective molecular fluid. A
  Mayer-cluster expansion for moments of the local static density
  fluctuations is set up. The simplest non-trivial contribution to this
  series leads back to mean-field theory. The central quantity of
  mean-field theory is the distribution of localization lengths, which
  we compute for all connectivities. The highly crosslinked gel is
  characterized by a one-to-one correspondence of connectivity and
  localization length. Taking into account higher contributions in the
  Mayer-cluster expansion, systematic corrections to mean-field can be
  included. The sol-gel transition shifts to a higher number of
  crosslinks per particle, as more compact structures are favored. The
  critical behavior of the model remains unchanged as long as finite
  truncations of the cluster expansion are considered.  To complete the
  picture, we also discuss various geometrical properties of the
  crosslink network, e.g. connectivity correlations, and relate the
  studied crosslink ensemble to a wider class of ensembles, including
  the Deam-Edwards distribution.
\end{abstract}


\maketitle

\section{Introduction}
\label{sec:intro}

In this paper we study chemical gelation, i.e. the equilibrium
transition from a liquid (sol) to an amorphous solid (gel) which is
induced by the introduction of permanent random crosslinks between the
particles of the liquid.  The classical theory of gelation, as
developed by Flory and Stockmayer \cite{FlSt}, assumes a tree like
connectivity of the random macromolecular networks generated in the
process of gelation and vulcanization. The critical exponents are
those of mean-field percolation. Stauffer \cite{St} and de Gennes
\cite{Ge} used instead the lattice connectivities of finite
dimensional percolation, embedding the gelation transition in the
context of critical phenomena. The resulting critical exponents are
those of three-dimensional percolation theory, in contradiction with
the classical values. Experimental support has accumulated for the
non-classical values \cite{Adam} and, in addition, given evidence for
the size of the critical region, outside of which mean-field exponents
prevail \cite{Ge2}.  One concludes that the geometric connectivity of gels 
is well described by percolation theory. On the other hand, the thermal
properties of gels are beyond the scope of percolation theory and have
in recent years been addressed with statistical mechanics, considering
both geometric and thermal fluctuations \cite{DeEd,BaEd,GoGo,Pa,PaRa}. A
mean-field picture has been developed \cite{GoZi,Castillo,CaGoZi} and a
renormalization group analysis \cite{Kane,Jansen} has been carried out
for the fluid side of the transition. Thereby the approach of
statistical mechanics has been connected to percolation
theory. However, the structural and elastic properties of
finite-dimensional gels are still only known within mean-field theory.

In this paper we focus on the {\it gel} phase for an ensemble of
crosslinks, which is given by continuum percolation. This is {\it not}
the Deam-Edwards distribution \cite{DeEd}, which has been used
frequently in the statistical approach. After a short introduction of
the model in Sec. \ref{sec:model}, we discuss geometrical properties
of the network, generated by three-dimensional percolation, e.g. we
investigate correlations in connectivity of neighboring sites and
compute the number of small loops
(Sec. \ref{sec:Xlinks}). Subsequently we set up a Mayer-cluster
expansion for the crosslinked melt (Sec. \ref{sec:expansion}). 
Our effort is not directed towards the critical behavior of the
gelation transition, which was studied in Refs. \cite{Kane,Jansen};
rather we want to find out, whether the characteristics of the gel
phase as described by mean-field theory, survive in a
finite-dimensional model. We recall that already on the level of a
mean-field theory one needs a distribution of length scales to 
characterize the local static density fluctuations of the gel
phase. Our aim is to find out in how far this picture has to be
modified in a short-range model.

The lowest diagram of the Mayer cluster expansion leads back to
mean-field theory (Sec. \ref{sec:truncations}), which is expected to
be appropriate deep in the gel phase away from the transition
point. We compute the distribution of localization lengths for all
connectivities. For the strongly connected gel we find a multi-peak
structure of the distribution, such that each coordination number
corresponds, to leading order, to a well-defined localization
length. Subsequently the first correction to mean-field theory within
the Mayer-cluster expansion is calculated. The critical connectivity
is increased due to the existence of small loops, which do not
increase the cluster size.  The gel fraction and the distribution of
localization lengths change only quantitatively, as compared to
mean-field theory and the critical exponents remain the same as long
as we only consider bare perturbation theory and do not resum the
Mayer-cluster expansion.  Finally in Sec. \ref{sec:deamedwards}, we
discuss a general class of crosslink distributions, which includes the
Deam-Edwards distribution as well as the distribution of
$d$-dimensional percolation.

\section{The model: crosslinked point particles}
\label{sec:model}

We consider a system of $N$ identical classical particles, confined to 
a $d$-dimensional volume $V$ with the average density of particles
$\rho_0=N/V$ being constant in the thermodynamic limit
$N,V\to\infty$. The particle positions are denoted by $R=\{\vec
R_i\}$, with $i$ running from 1 to $N$.  The interactions are given by 
the Hamiltonian
\begin{equation}
  \label{eq:hamiltonian}
  H(R) = U(R) + \sum_{e=1}^M V(\vec R_{i_e}-\vec R_{j_e})\ .
\end{equation}
Here, $U(R)$ describes the particle interactions in the fluid, without
crosslinks, and is given by a sum over single-particle and pair
potentials. Permanent random crosslinks are introduced between $M$
pairs of particles, numbered by $\{(i_e,j_e)\}_{e=1}^M$. Two monomers,
participating in a crosslink, are forced to remain close to each
other. We model this constraint by an attractive pair potential
$V(\vec R_{i_e}-\vec R_{j_e})$. The simplest choice is a harmonic one, 
$V(\vec R_{i_e}-\vec R_{j_e}) = \frac \kappa 2 (\vec R_{i_e}-\vec
R_{j_e})^2$, where $\kappa$ is the strength of the
crosslink coupling. It has been shown that the
harmonic potential is equivalent to crosslinks represented by hard
constraints in the limit $\kappa \to \infty$ \cite{Vilgis}. Note that
further geometrical constraints arise indirectly via the interlocking 
of closed loops of crosslinks, see the discussion in \cite{CaGoZi}. 
These effects are not taken into account in the above Hamiltonian.

It is convenient to introduce the symmetric $N\times N$ connectivity 
matrix $J$ with entries
\begin{equation}
J_{ij} = \sum_{e=1}^M (\delta_{i,i_e}\delta_{j,j_e}+
\delta_{i,j_e}\delta_{j,i_e})\ .
\end{equation}
These entries equal one, whenever $i$ and $j$ are linked together, 
and zero otherwise. Multiple crosslinks are excluded. Harmonic
interactions due to the crosslinks can thus be represented by
\begin{equation}
\sum_{e=1}^M V(\vec R_{i_e}-\vec R_{j_e}) 
= \frac{\kappa}{2} \sum_{1\leq i<j\leq N} \; J_{ij} (\vec R_i-\vec R_j)^2\ .
\end{equation}
The thermal degrees of freedom are the positions of the monomers. The
set of crosslinks ${\cal C}:=\{(i_e,j_e)\}_{e=1}^M$ represents the
quenched disorder of the model, so that the monomers equilibrate in the
presence of a fixed, non-equilibrium configuration of crosslinks.

All thermodynamic properties can be obtained from the partition
function 
\begin{equation}
  \label{eq:Z1}
  Z({\cal C}) = \int_{V^N} d^{dN} R \  \exp\left\{ - \beta H \right\} \ . 
\end{equation}
which still depends on the quenched disorder ${\cal C}$.  Here 
$\beta=1/(k_B T)$ denotes the inverse temperature. One usually assumes
that the Gibbs free energy, $\beta F=-\log Z$, is self-averaging in the
thermodynamic limit, and thus computes its average over all crosslink
realizations, ${\overline F}$.

\section{Distribution of Crosslinks: Finite-dimensional percolation}
\label{sec:Xlinks}

In this section, we give an explicit formula for the distribution of
crosslinks corresponding to $d$-dimensional percolation. It is based on
the intuitive picture that a given number of crosslinks $M$ is
introduced simultaneously and instantaneously into the fluid of
monomers. The crosslinks are strictly bivalent and connect pairs of
monomers by a chemical bond. Given the instantaneous (or more
realistically fast) reaction, pairs of monomers which are nearby in
the instant of crosslinking, have a high probability to be connected,
whereas pairs of monomers which are distant, have a very small
probability to be crosslinked. The assumption of a fast crosslinking
reaction as compared to the diffusive time-scale of the fluid
molecules is realistic and becomes better and better the closer one
gets to the gelation transition, because larger and larger clusters
are built up and give rise to increasingly longer relaxation times of
the molecules in the fluid.

The distribution is generated in two steps:

1) A liquid configuration $R^0=\{\vec R_i^0\}_{i=1}^N$ (an instant) of
a $d$-dimensional fluid is generated, by randomly choosing $N$ points in
$d$-dimensional space. One possibility  is to generate liquid
configurations with the Boltzmann weight of the uncrosslinked system,
thereby including short-range correlations of the fluid.
Such a procedure will be discussed in Sec. \ref{sec:deamedwards}. 
Here we consider the simpler case of random, uncorrelated positions
$R^0=\{\vec R_i^0\}_{i=1}^N$ in order to model crosslinks
corresponding to $d$-dimensional percolation.

2) Given the configuration $R^0$, each crosslink is chosen
independently, as described by a factorized distribution
\begin{equation}
\label{eq:disorder2}
P({\cal C}|R^0)=\prod_{e=1}^M p(i_e,j_e|R^0).
\end{equation}
The probability for a particular crosslink $(i_e,j_e)$ 
depends only on the relative distance $|\vec R^0_{i_e j_e}|$
\begin{equation}
\label{rdependence}
p(i_e,j_e|R^0)=\frac{\Delta(|\vec R^0_{i_e j_e}|)}
{\sum_{i,j=1}^N\Delta(|\vec R^0_{ij}|)}
\end{equation}
with $\vec R_{ij}^0 = \vec R_i^0 - \vec R_j^0$.
The function $\Delta(x)$ should be of finite range, examples are 
$\Delta(x)=\theta(\lambda-x)$ or $\Delta(x)=e^{-x/\lambda}$. The
denominator in Eq. (\ref{rdependence}) ensures the proper normalization.
The average of an observable $f(R^0,{\cal C})$ over all crosslink
configurations ${\cal C}$ is given by
\begin{equation}
\label{average}
\overline{f(R^0,{\cal C})}= \int_{V^N} \frac{d^{dN} R^0}{V^N} \quad
\sum_{i_1,j_1=1}^N \cdots \sum_{i_M,j_M=1}^N \,
P({\cal C}|R^0)\ f(R^0,{\cal C})\ .
\end{equation}
Note that the integration over $R^0$ is part of the averaging
over all crosslink distributions and should not be confused with
thermal averages.

Sometimes, it is technically simpler to allow the total number of
crosslinks to fluctuate. To this end, we replace the second step in
the above procedure by the following: 

2') Given the configuration $R^0$, choose each 
crosslink independently, as described by a factorized distribution
\begin{equation}
P(\{J_{ij}\}_{i<j}|R^0)=\prod_{1\leq i<j\leq N}
\left(\delta(J_{ij}-1)p(\vec R^0_{ij})
+\delta(J_{ij})(1-p(\vec R^0_{ij})) 
\right)
\end{equation}
which explicitely excludes multiple crosslinks.
The function $p(\vec x)$ should take values $0\leq p(\vec x) \leq 1$ 
and be of finite range. It will be convenient to assume a Gaussian 
shape
\begin{equation}
  \label{eq:probability}
  p(\vec R_{ij}^0) =
  \exp\left\{-\frac{a}{2}(\vec R_{ij}^0)^2\right\}\ .
\end{equation}
The ``crosslinking-length'' $\ell=a^{-1/2}$ is chosen comparable 
to the mean distance between particles, $\rho_0^{-1/d}$, in order to
guarantee an extensive number of crosslinks (see the following
section). The average (\ref{average}) of an observable over all 
crosslink configurations, now equivalently denoted by ${\cal C}:=
\{J_{ij}\}_{i<j}$, is then replaced by
\begin{equation}
\overline{f(R^0,{\cal C})}=\int_{V^N} \frac{d^{dN} R^0}{V^N} \quad
\int \prod_{i<j}d J_{ij}\ P({\cal C}|R^0)\ f(R^0,{\cal C})\ .
\end{equation}

Before discussing the physics of the crosslinked system, we will
investigate some geometrical properties of the network of crosslinks.
Special emphasis is given to local structures which are characteristic for
the low-dimensional structure of the network. They differ
substantially from the properties of diluted random graphs \cite{Bo} 
which give the proper mean-field description of the disorder 
distribution.

\subsection{Number of crosslinks}

To start with, we compute the distribution of the total number of
crosslinks. As each pair of monomers is considered independently in
the crosslinking process, the crosslink number is expected to fluctuate
according to a Poissonian distribution. For the probability $W(M)$ of
having $M$ crosslinks we indeed find
\begin{eqnarray}
  \label{eq:Mdistribution}
  W(M) &=& \overline{ \delta_{M,\sum_{i<j} J_{ij}}}
  \non\\
  &=& \int_{-\pi}^\pi \frac{dx}{2\pi} 
  \overline{ e^{ix(M-\sum_{i<j} J_{ij})}} \non\\
  &=& \int_{-\pi}^\pi \frac{dx}{2\pi} e^{ixM}   
  \left(1-\frac{W_-}{V}+ \frac{W_-}{V} e^{-ix} \right)^{N \choose 2}
  \non\\ 
  &=& {\frac{N(N-1)}{2} \choose M} \left(1-\frac{W_-}{V}
  \right)^{\frac{N(N-1)}{2}-M }
  \left(\frac{W_-}{V} \right)^M\ ,
\end{eqnarray}
where $W_- := \int_V d^d R\ p(\vec R) = (2\pi/a)^{d/2}$ denotes 
the effective crosslinking volume for the distribution of Eq.\ 
(\ref{eq:probability}). In the limit of large $N$, with constant 
particle density $\rho_0=N/V$, these probabilities tend to a
Poissonian 
\begin{equation}
  \label{eq:poissonM}
  W(M) \to  \frac{\overline M ^M}{M!}  e^{-\overline M}
\end{equation}
with mean $\overline M = \rho_0 W_- N/2$. Fluctuations around the mean
are of order ${\cal O}(\sqrt{\overline{M^2}-\overline M^2})
={\cal O}(\sqrt{\overline M})$,
and hence small compared to $\overline M$. We therefore expect the
differences between the ensembles with fixed and fluctuating number of
crosslinks to disappear in the thermodynamic limit.

\subsection{Distribution of coordination numbers}

The simplest local property of the crosslink network is given by the 
distribution of coordination numbers. The average fraction of
particles being connected to exactly $k$ other particles is given by
\begin{eqnarray}
  \label{eq:connect}
  w(k) &=& \overline{ \frac{1}{N} \sum_i \delta_{k,\sum_j J_{ij}}}
  \non\\
  &=& \overline{ \delta_{k,\sum_j J_{ij}}} \non\\
  &=& \int_{-\pi}^\pi \frac{dx}{2\pi} 
  \overline{ e^{ix(k-\sum_j J_{ij})}} \non\\
  &=& \int_{-\pi}^\pi \frac{dx}{2\pi} e^{ikx}   
  \left(1-\frac{W_-}{V}+ \frac{W_-}{V} e^{-ix} \right)^{N-1} \non\\
  &=& {N \choose k} \left(1-\frac{W_-}{V} \right)^{N-k}
  \left(\frac{W_-}{V} \right)^k \non\\
  &\to&  e^{-c} \frac{c^k}{k!}\ .
\end{eqnarray}
where the last expression describes the limiting distribution for
$N\to\infty$. It is thus found to equal a Poissonian of mean 
$c :=\rho_0 W_-$.
The same coordinations are found in the long-ranged case $p(\vec
R_{ij}^0) = c/N$. Differences to the long-ranged case show up
in nonlocal features, as connectivity correlations and occurrence of
loops.

\subsection{Connectivity correlations}
\label{sec:conncorr}

To study connectivity correlations, we consider the coordination
numbers of particles which are directly connected by a single
crosslink. Given a large network, we determine the (asymptotic)
probability that a randomly and uniformly selected crosslink has
endpoints of coordinations $k_1$ and $k_2$ respectively. This quantity
is given by the sum over all links, which have one endpoint connected
to $k_1$ and the other endpoint connected to $k_2$ particles,
normalized by the total number of crosslinks
\begin{eqnarray}
  \label{eq:conncorr1}
  w(k_1,k_2) &=& \frac{2}{\rho_0 W_- N} \sum_{i<j} J_{ij}\ 
  \delta_{k_1, \sum_l J_{il}}\ \delta_{k_2, \sum_l J_{jl}}\non\\
  &=& \frac{2}{\rho_0 W_- N} \sum_{i<j} J_{ij}\ 
  \delta_{k_1-1, \sum_{l\neq i,j} J_{il}}\ 
  \delta_{k_2-1, \sum_{l\neq i,j} J_{jl}}\non\\
  &=& \frac{N-1}{\rho_0 W_-} \int_{-\pi}^\pi \frac{dx_1\ dx_2}{4\pi^2}
  e^{i x_1(k_1-1) + i x_2(k_2-1)} \overline{ J_{12} e^{-i \sum_{l\geq
        3} (J_{1l}x_1+J_{2l}x_2)}}\ .
\end{eqnarray}
The disorder distribution of Eq.(\ref{eq:probability}) allows to
compute the average up to summations
\begin{eqnarray}
  \label{eq:conncorr}
  w(k_1,k_2) &=& \sum_{n=0}^{\min(k_1,k_2)-1} \sum_{l=0}^\infty
  \sum_{m=0}^{k_1+k_2-2-2n} { k_1+k_2-2-2n \choose m }
  \frac{(-1)^m\ c^{k_1+k_2-2-n+l}\ e^{-2c}}{
  n!\ (k_1-1-n)!\ (k_2-1-n)!\ l!} \non\\
  && \ \ \ \ \ \times \frac{1}{2^{d(n+l+m-1)/2}\ (2+n+l+m)^{d/2}}\ , 
\end{eqnarray}
where $c=\rho_0 W_-$ denotes again the average connectivity.

\begin{figure}[htb]
\begin{center}
\myscalebox{\includegraphics{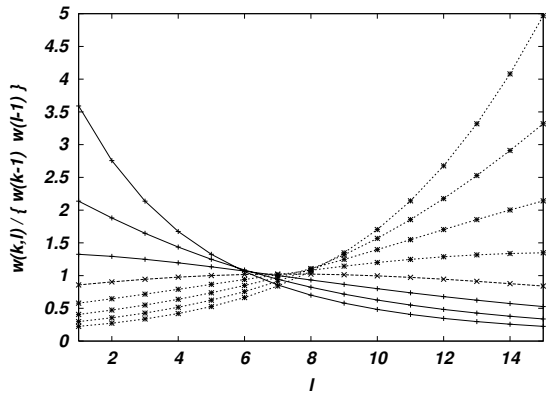}}
\end{center}
\caption{Connectivity correlations for $d=3$ and average connectivity 
$c=\rho_0 W_- = 6$. The figure shows the probability that a randomly
selected crosslink has endpoints of connectivities $k$ resp. $l$, 
as  a function of $l$ for various values of $k=1,3,5,7,9,11,13,15$ (from
top to bottom on left axis, lines are guides to the eyes), and 
normalized by the uncorrelated probability $w(k-1) w(l-1)$. 
Values larger than one correspond to positive correlations, values 
smaller than one to anti-correlations.}
\label{fig:conncorr}
\end{figure}

The above expression has been evaluated numerically and is represented
in Fig. \ref{fig:conncorr} for $d=3$ and $c=6$. There exist obviously
connectivity correlations in the sense that particles of low
coordination are surrounded more likely by other low-connected
particles. Similarly, particles of high coordination are likely to
be surrounded by other high-coordinated particles. This effect
can be understood intuitively: Low coordination of a particle $i$
corresponds to a small number of other particles inside the effective
crosslinking volume $W_-$ centered in $\vec R_i^0$. Due to the overlap
of this volume with the crosslinking volume of the neighboring
particles, these will typically have a small number of potential
crosslinking partners, too.  The curves in Fig. \ref{fig:conncorr}
cross 1 in the vicinity of $k_2\simeq c +1$, which is the average
connectivity of particles found by considering endpoints of randomly
chosen crosslinks.

Connectivity correlations decrease exponentially with spatial
dimension $d$ and vanish in the limit $d\to \infty$. In this limit,
the crosslink network becomes a random graph in the sense of
\cite{Erdoes}.

\subsection{Small loops}

Another signature of the finite-dimensional structure of the
crosslink network is the existence of an extensive number of short
loops and other local, non-treelike sub-graphs. For example, the mean  
number of triangles is easily computed,
\begin{eqnarray}
  \label{eq:triangles}
  {\cal N}_\Delta &=& \sum_{i<j<k} \overline{J_{ij} J_{jk} J_{ki}}  
  \non\\
  &=& {N\choose 3} \overline{ J_{12} J_{23} J_{13}} \non\\
  &=& {N\choose 3} \int \frac{ d^d R_1^0\ d^d R_2^0\ d^d R_3^0
    }{ V^3 }\ p(\vec R_1^0-\vec R_2^0)\ p(\vec R_2^0-\vec R_3^0)\ 
   p(\vec R_1^0-\vec R_3^0) \non\\
  &=& \frac{N}{6} \rho_0^2 \int  d^d R_1^0\ d^d R_2^0\
   p(\vec R_1^0)\ p(\vec R_2^0)\ p(\vec R_1^0-\vec R_2^0) \non\\
  &=:& \frac{N}{6} \rho_0^2 W_\Delta\ .
\end{eqnarray}
and seen to be {\it extensive} for generic short ranged distributions
$p(\vec R)$. For the particular choice of Eq.\ (\ref{eq:probability})
we find $\rho_0^2 W_\Delta=\rho_0^2 (2\pi/\sqrt{3}a)^d = c^2 3^{-d/2}$.
For constant average connectivity, the number of triangles tends to
zero as $d\to\infty$. In this limit we recover the properties of
random graphs, which are known to be locally tree-like.

The presented derivation can be easily generalized to more complicated
local structures, always giving rise to extensive numbers decreasing
exponentially with growing spatial dimension.

\section{Mayer-cluster expansion of the replicated local density
  function}
\label{sec:expansion}

After having discussed some geometrical properties of the random
networks generated according to finite-dimensional percolation, we go
back to the original physical model of a crosslinked fluid, as given
by Eq.\ (\ref{eq:hamiltonian}). We discuss the order parameter of the
gel phase and show how to compute local static density fluctuations
within a Mayer-cluster expansion. 

\subsection{Localization of particles and the physical order
  parameter}
\label{sec:physicalop}

As discussed previously \cite{GoGo,GoZi,CaGoZi}, the sol-gel
transition is an equilibrium phase transition from a liquid state
(sol) to an amorphous solid state (gel). In the gel a finite fraction
of particles is localized in the vicinity of fixed equilibrium
positions. Increasing the number of crosslinks, localization first
occurs, when a macroscopic cluster of crosslinked molecules appears,
i.e. at the percolation threshold.  Hence the fraction of localized
particles is determined by the mass of the macroscopic cluster.

Due to the random crosslinks, the equilibrium positions of the
localized particles do not exhibit any periodic structure, but are
random.  The gel is an amorphous solid, but still all macroscopic
properties have to be translationally invariant.  In particular, the
single-particle density
\begin{equation}
  \label{eq:rho1}
  \rho^{(1)}(\vec r) = \sum_{i=1}^N \overline{
  \left\langle \delta(\vec r -\vec R_i) \right\rangle} = \rho_0
\end{equation}
is homogeneous in both phases. Here and in the following, the thermal
average with respect to the equilibrium state of the crosslinked
system, cf. Eq. (\ref{eq:Z1}), is denoted by $\langle \cdot \rangle$.

To detect localization, one has to consider higher moments of the
local density. The simplest one is  
\begin{equation}
  \label{eq:rho2}
  \rho^{(2)}(\vec r\,^1, \vec r\,^2) = \sum_{i=1}^N \overline{
  \left\langle \delta( \vec r\,^1 -\vec R_i) \right\rangle 
  \left\langle \delta( \vec r\,^2 -\vec R_i) \right\rangle }\ .
\end{equation}
Macroscopic homogeneity implies that $\rho^{(2)}= \rho^{(2)}(\vec
r\,^1- \vec r\,^2)$. In the sol phase, all particles are free to
explore the whole container and hence any particular one is equally
likely to be in any sub-volume. This implies $\langle \delta( \vec
r\,^1 -\vec R_i) \rangle =1/V $ and hence $\rho^{(2)}= \rho^{(2)}(\vec
r\,^1- \vec r\,^2)=\rho_0/V$. In the gel phase, however, a finite
fraction of all particles is part of the macroscopic cluster and thus
localized. In the simplest model \cite{Castillo}, one assumes Gaussian
localization $\langle \delta( \vec r\,^1 -\vec R_i) \rangle \propto 
\exp\{-(\vec r\,^1-\vec a_i)^2/(2\xi_i^2)\} $ around homogeneously
distributed random localization centers $\vec a_i$. Here
$\xi_i$ is the localization length, characterizing the extent of thermal
fluctuations around the preferred position $\vec a_i$. 
The second moment is then given by
\begin{eqnarray}
\label{eq:Gaussloc}
  \rho^{(2)}(\vec r\,^1- \vec r\,^2)& =& \overline{\sum_{i=1}^N \int 
\frac{d^d a_i}{V} \left( \frac 1{2\pi\xi_i^2}\right)^d
\exp\left\{-\frac1{2\xi_i^2}(\vec r\,^1-\vec a_i)^2
        -\frac1{2\xi_i^2}(\vec r\,^2-\vec a_i)^2\right\} }\nonumber \\
&=&\rho_0 \int_0^\infty d \xi^2 \tilde P(\xi^2)  \left( \frac
  1{4\pi\xi^2}\right)^{\frac d2}
  \exp\left\{- \frac1{4\xi^2} (\vec r\,^1-\vec r\,^2)^2 \right\}
\end{eqnarray}
The second moment is thus expressed in terms of the distribution of
localization lengths
\begin{equation}
\tilde P(\xi^2):=\frac{1}{N}\sum_{i=1}^N
\overline{\delta(\xi^2-\xi_i^2)} 
\end{equation}
 which is expected to be rather broad due to the
inhomogeneous environment of different particles: Some
particles are expected to be strongly localized due to high local
connectivity (steep local potentials), whereas particles e.g. on
dangling bonds are expected to exhibit larger spatial fluctuations,
corresponding to larger localization lengths.

The gelation transition occurs, when the connectivity per particle is
${\cal O}(1)$. Hence, there is no reason to assume Gaussian static
density fluctuations, and the second moment (\ref{eq:rho2}) is not
sufficient to completely characterize the structure of the gel
phase. Instead, the full distribution of local static density
fluctuations is required, or, equivalently, all moments
\begin{equation}
\label{eq:rhol1}
  \rho^{(l)}(\vec r\,^1,... \vec r\,^l) = \sum_{i=1}^N\overline{ 
  \prod_{j=1}^l \left\langle \delta( \vec r\,^j -\vec R_i) 
  \right\rangle}\ .
\end{equation}
Within mean-field theory, it has been shown that
all higher moments can be expressed in terms of $P(\xi^2)$, according to
\begin{equation}
\label{eq:rhol2}
  \rho^{(l)}(\vec r\,^1,... \vec r\,^l) = \rho_0 \int d^d a 
  \int_0^\infty d\xi^2 \tilde P(\xi^2)
  \left(\frac1{2\pi\xi^2}\right)^{\frac{dl}2} 
  \exp\left\{- \frac 1{2\xi^2} \sum_{j=1}^l (\vec r\,^j-\vec a)^2
  \right\} 
\end{equation}
so that the structure is characterized by a single function, the
distribution of localization lengths. Whether or not this result holds
beyond mean-field theory is an open question.

\subsection{From the replicated partition function to an effective
  molecular fluid}

Our aim is the computation of the partition function of
Eq.\ (\ref{eq:Z1}). The most important pair potential $U$ is the
excluded volume interaction. It is known that the crosslinked
melt without excluded volume will collapse. Hence we cannot simply 
set $U(R)=0$. Here we introduce instead a Lagrange parameter to
ensure the constraint of homogeneous density. To this end we choose 
\begin{equation}
\label{eq:U}
-\beta U(R) =\int_{V^N} d^{dN}x\ \mu(\vec x)
\sum_i \delta(\vec x - \vec R_i)=
\sum_i \mu(\vec R_i)
\end{equation}
and consider $\mu(\vec R)$ as a Lagrangian multiplier coupled to the
single-particle density. It will be determined such that the single
particle density remains homogeneous. Technically it is much simpler
to work with single-particle potentials instead of the excluded volume
interaction.

We expect the model to be self-averaging in the macroscopic limit,
{\it i.e.} intensive observables should not depend on the particular
disorder realization, but only on the statistical ensemble of
crosslinks. Hence we set out to compute the disorder averaged free
energy $F:=-\overline{\ln Z({\cal C})}$ with
\begin{equation}
  \label{eq:Z}
  Z({\cal C}) = \int_{V^N} d^{dN} R \  \exp\left\{ \sum_i \mu(\vec
    R_i)-  \sum_{i<j} J_{ij} V(\vec R_i-\vec R_j) \right\} \ . 
\end{equation}
We use units of energy such that $\beta=1$. 

To perform the average over the crosslink distribution we use the
replica trick \cite{Edwards}
\begin{equation}
  \label{eq:replicatrick}
  \overline{\ln Z({\cal C})} = \lim_{n\to 0} \frac{
    \overline{Z({\cal C})^n} -1}{n}\ , 
\end{equation}
first assuming positive integer $n$, and using a replica-symmetric
ansatz to analytically continue the results to $n\to 0$ at the
end. For integer $n$, the model is replaced by $n$ copies $R^a,
a=1,...,n,$ with independent coordinates but identical disorder.
We can explicitly calculate the average over the $J_{ij}$, for given
$R^0$, and find
\begin{eqnarray}
  \label{eq:Zn}
  \overline{Z({\cal C})^n} &=& \int_{V^{nN}} \prod_{a=1}^n d^{dN} 
  R^a \ \overline{ \exp\left\{  \sum_i\sum_{a=1}^n \mu(\vec R_i^a)- 
  \sum_{i<j} J_{ij} 
  \sum_{a=1}^n V(\vec R_i^a-\vec R_j^a)\right\} } \non\\
  &=& \frac1{V^N} \int_{V^{(n+1)N}} \prod_{a=0}^n d^{dN} R^a 
   \exp\left\{ \sum_i\sum_{a=1}^n \mu(\vec R_i^a)+ 
    \sum_{i<j} \ln \left[ 1+p(\vec R_i^0 - \vec R_j^0)
    \left( e^{-\sum_{a=1}^n V(\vec R_i^a - \vec R_j^a)} -1
    \right) \right]  \right\}\ .
\end{eqnarray}
Note that the integration in the last line runs also over all
disorder configurations $\vec R_i^0$. It thus resembles an
$(n+1)$-times replicated system, where the replica of index 0
corresponds to the liquid configuration being part of the quenched 
disorder, and the replicas $a=1,...,n$ are the thermal degrees of 
freedom of the crosslinked model.

The main idea is now to interpret this expression as the partition
function of a fluid of $N$ ``effective molecules'', each consisting 
of $n+1$ particles, cf. \cite{Go}.  We simplify the notation by 
introducing a $d(n+1)$-dimensional vector 
$\hat R_i = (\vec R_i^0,...,\vec R_i^n)$
for the position vectors of the ``constituents'' of molecule $i$. 
The molecules interact pairwise via the potential
\begin{equation}
\label{eq:WWmolecule}
V_{n}(\hat R_i-\hat R_j) = -\ln \left[1+p(\vec R_i^0 - \vec R_j^0)
 \left( e^{-\sum_{a=1}^n V(\vec R_i^a-\vec R_j^a)}-1 \right)\right]\ ,
\end{equation} 
which is symmetric with respect to permutations of the last $n$
particles, thus reflecting the replica symmetry of the replicated
partition function. Note that there are no explicit {\it
intra}-molecular interactions. However, the {\it inter}-molecular
interactions are many-particle interactions, thus leading to an
effective coupling of different particles within one molecule.  In the
context of gelation, the central question is whether or not molecules
are bound in the sense, that $|\vec R_i^a-\vec R_i^b|,\ a\neq b,$
stays finite for a finite fraction of all molecules $i$ in the
thermodynamic limit.  As we will see in the following, unbound
molecules may be identified with the sol fraction, bound molecules
with the gel fraction which appears only at or above the gelation
transition.

\subsection{Mayer-cluster expansion for the effective molecular fluid}

Having in mind the interpretation of the replicated and (partially)
disorder-averaged system as an effective fluid, we can use the
classical concepts of equilibrium statistical mechanics of fluids
\cite{Cr}, in particular the Mayer-cluster expansion 
in the diagrammatic formulation of \cite{MoHi}. 
We thus introduce the Mayer-bond
\begin{eqnarray}
  \label{eq:Mayerbond}
  b(\hat R_i,\hat R_j) &=& e^{-V_{n}(\hat R_i-\hat R_j)} -1 \non\\
  &=& p(\vec R_i^0 -\vec R_j^0) \left[ e^{-\sum_{a=1}^n 
  V(\vec R_i^a -\vec R_j^a)} -1 \right] \ ,
\end{eqnarray}
which factorizes into a $n$-fold replicated Mayer-bond of a fully
connected system times the probability of existence of the crosslink 
$J_{ij}=1$.

The quantity of central interest for gelation are the moments of the
local density (\ref{eq:rhol1}), all of which can be related to the 
single-{\it molecule} density
\begin{equation}
  \label{eq:density}
  \rho(\hat r) =  \sum_{i=1}^N \left\langle \delta(\hat r - \hat R_i)
  \right\rangle_n \ .
\end{equation}
The brackets $\langle\cdot\rangle_n$ denote the thermodynamic average 
of the effective molecular fluid, cf. Eq. (\ref{eq:Zn}). 
The moments of the local density $\rho^{(l)}(\vec r\,^1,... \vec r\,^l)$
are obtained from the molecule density according to
\begin{equation}
  \label{eq:physop}
  \rho^{(l)}(\vec r\,^1,...,\vec r\,^l) = \lim_{n\to 0} \int_{V} 
  d^d r^0 \prod_{a=l+1}^n \int_V d^d r^a \ \rho(\hat r)\ .
\end{equation}
For a
macroscopically translationally invariant system, as the gel, 
all single-{\it particle} densities have to be homogeneous, implying 
for the molecule density
\begin{equation}
  \label{eq:homogen}
  \lim_{n\to 0} \int_{V^n} d^d r^0\cdots d^d r^{a-1} \ d^d 
  r^{a+1} \cdots d^d r^n \ \rho(\hat r) = \rho_0
\end{equation}
for all $a=0,...,n$, cf. Eq. (\ref{eq:rho1}).

\begin{figure}[htb]
\begin{center}
\myscalebox{\includegraphics{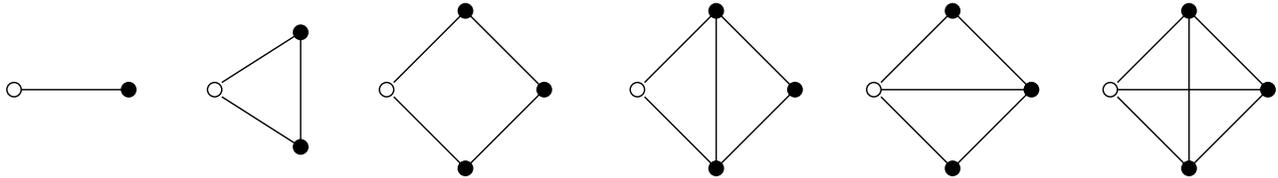}}
\end{center}
\caption{Simplest diagrams in the Mayer-cluster expansion of the
local replicated density. These diagrams are not reducible to two
disconnected components by deleting a single vertex.}
\label{fig:diagram}
\end{figure}

The following can be understood best by using a graphical
representation. We consider diagrams whose vertices are either
white circles, representing a molecule at position $\hat r$, or
black circles, representing an integration over the single-molecule
density. The lines connecting two circles denote a Mayer-bond
$b(\hat r_1,\hat r_2)$. Each diagram has to be divided by its
symmetry number which counts the number of possible permutations
of vertices which do not alter the diagram. Simple examples are
given in Fig. \ref{fig:diagram}, {\it e.g.} the first diagram
reads $\int  d^{d(n+1)} \hat r_1 b( \hat r, \hat r_1) \rho(\hat r_1)$.
The symmetry number equals 1 as black and white circles are
distinguishable. Two circles are never connected by more than one
Mayer-bond. A diagram will be called {\it one-vertex irreducible},
if it does not reduce to two disconnected diagrams by deleting any
single vertex and its adjacent Mayer-bonds. All diagrams in Fig.
\ref{fig:diagram} are one-vertex irreducible.

Using the results of \cite{MoHi}, $\rho(\hat r)$ has to fulfill
the non-linear integral equation
\begin{eqnarray}
  \label{eq:expansion}
\ln \rho(\hat r) = \mu(\hat r) + &\sum& 
  \mbox{ one-vertex irreducible
  diagrams with one white circle of coordinate $\hat r$,} \non\\
  && \mbox{ and an arbitrary number of black circles}\ .
\end{eqnarray}
As already mentioned above, the single-particle potentials $\mu(\hat
r)= \sum_a \mu(\vec r\,^a)$ act as Lagrangian multipliers guarantying
the homogeneity condition (\ref{eq:homogen}). The first diagrams
with up to four vertices are shown in Fig. \ref{fig:diagram}.

\subsection{Structure of the density: Distribution of localization
  lengths}
\label{sec:op}

In order to extract the moments of the local density according to
Eq. (\ref{eq:physop}), we have to perform the replica limit $n\to
0$. We therefore introduce some ansatz on the
analytical structure of $\rho(\hat r)$. The first assumption concerns
the validity of replica symmetry at the order-parameter level: We
assume that $\rho(\hat r)$ is invariant under any permutation of
the $n$ replicas $a=1,...,n$ of the crosslinked system. Only the
disorder system $a=0$ is distinct. The second assumption is inspired 
by the simple model of Gaussian localization of particles belonging to 
the gel. Following Eq. (\ref{eq:rhol2}) we represent $\rho(\hat r)$ 
by
\begin{equation}
  \label{eq:rs}
  \rho(\hat r) = (1-q)\frac{\rho_0}{V^n} +
  q \rho_0  \int d^d R \int_0^\infty d\tau_0 \ d \tau\ 
  P(\tau_0, \tau)\ \left( \frac{\tau_0}{2\pi} \right) 
  ^\frac{d}{2} \left( \frac{\tau}{2\pi} 
  \right)^\frac{dn}{2} \exp\left\{ -\frac{\tau_0}{2}(\vec r\,^0
  - \vec R)^2 - \frac{\tau}{2} \sum_{a=1}^n (\vec r\,^a-\vec R)^2
  \right\}\ .
\end{equation}
The interpretation of this ansatz is quite simple: A fraction $1-q$ of
effective molecules, with $0\leq q \leq 1$, is not bound, the $n+1$
particles of each molecule are homogeneously distributed in V. The
remaining $qN$ molecules are restricted by the distribution of inverse
squared localization lengths $P(\tau_0,\tau)$. Given a particular 
$\vec r\,^0$ from the disorder distribution, we find a randomly drawn
$\vec R$ at an average squared distance $1/\tau_0$, which itself is
the localization center of the other $n$ particles. Non-zero $\tau$
and $\tau_0$ thus correspond to bound molecules, or, more physically,
to a situation where the particles are localized close to the original
disorder configuration which was drawn before crosslinking. Thus, $q$
denotes the gel fraction of the crosslinked material, {\it i.e.} the
fraction of particles which are part of its solid component. The
localization length $\xi^2$ of the phenomenological model
(Sec.\ \ref{sec:physicalop}) is simply related to $\tau=1/\xi^2$. Our
ansatz generalizes the ansatz introduced in \cite{Castillo} for the
Deam-Edwards distribution, with reduced permutation symmetry. Please
note that it already implies a homogeneous single-particle
density. Only the global normalization has to be enforced, thus we can
fix $\mu(\hat r)=\mu$ to be a constant chemical potential.

Ansatz (\ref{eq:rs}) in fact solves the Mayer-cluster expansion
(\ref{eq:expansion}) for an appropriate choice of $P(\tau_0,\tau)$.
This can be seen best by investigating the structure of the diagrams
on the left-hand side of Eq. (\ref{eq:expansion}). Due to the Gaussian
shape of the Mayer bonds (\ref{eq:Mayerbond}) all integrations
corresponding to black circles can be carried out, see Sec.
\ref{sec:truncations} for specific examples. The resulting expression
is again a (continuous) sum over Gaussian terms in the distances
$(\vec r\,^a -\vec R)$. Also the symmetry of ansatz (\ref{eq:rs}) is
preserved, because the Mayer-bonds are symmetric with respect to
permutations of the thermal replicas $a=1,...,n$, too. The
Mayer-cluster expansion thus leads, via comparing the coefficients of
the Gaussian contributions of identical variance on both sides of Eq. 
(\ref{eq:expansion}), to a non-linear integral equation for 
$P(\tau_0,\tau)$.
 
To calculate the physical order parameter, we have to plug the
replica-symmetric ansatz (\ref{eq:rs}) into
Eq. (\ref{eq:physop}). This can be done for arbitrary moments of the
local density. For the sake of clarity, we restrict the presentation
to the second moment ($l=2$), and find by direct integration over
$\vec r\,^0,\vec r\,^3,...,\vec r\,^n$ and $\vec R$
\begin{equation}
  \label{eq:physloc}
  \rho^{(2)}(\vec r\,^1,\vec r\,^2) =  (1-q) \frac{\rho_0}V + q \rho_0
  \int_0 ^\infty d \tau\ P(\tau) \left( \frac{\tau}{4\pi}\right) 
  ^\frac{d}{2} \exp\left\{ -\frac{\tau}{4}(\vec r\,^1
  - \vec r\,^2)^2 \right\}\ .
\end{equation}
The second moment depends on $P(\tau_0,\tau)$ only 
through the reduced distribution 
\begin{equation}
  \label{eq:poftau}
  P(\tau) := \int_0^\infty d\tau_0\ P(\tau_0,\tau)
\end{equation}
of localization lengths of the thermal system. The same is true for
all higher moments, implying that the information on
$\tau_0$ is not needed in the description of a gel. We therefore
concentrate our attention on the properties of the reduced
distribution $P(\tau)$ alone.

\section{Truncations of the Mayer-cluster expansion}
\label{sec:truncations}

Based on the Mayer-cluster expansion, one could try to partially resum
the series, developing {\it e.g.} a hypernetted-chain or Percus-Yevick
approximation \cite{Cr}. Due to unsolved problems in calculating the
replica limit of general diagrams, this seems, however, not to be
tractable. Furthermore the above summations are notoriously bad, as
far as critical exponents are concerned.  We therefore restrict our
investigations to the simplest truncations of the series. As we will
see in the following section, already the first diagram, consisting of
one white, one black vertex and a single Mayer-bond, reproduces the
results of mean-field theory.

In Sec. \ref{sec:triangle} we include the first correction:
a triangular diagram with one white and two black circles, being
completely connected by three Mayer-bonds. This diagram is shown as
the second one in Fig. \ref{fig:diagram}. We will show, that the
inclusion shifts the sol-gel transition to higher connectivities.
The critical properties remain unchanged; we find the same critical 
exponents for the growth of the gel fraction and for the 
scaling function of the inverse squared localization lengths.

{\it All} diagrams containing more than one Mayer-bond decrease
exponentially with the space-dimension $d$, making truncations more
reliable in high dimensions. However, the Mayer-cluster expansion
cannot be considered as a systematic expansion around the
infinite-dimensional mean-field.

\subsection{Back to mean-field theory}
\label{sec:meanfield}

We start with the simplest non-trivial truncation of the
Mayer-cluster expansion: only the leftmost diagram of Fig.
\ref{fig:diagram} will be included. Its contribution is denoted by
$\Gamma_{-}$ and is explicitly given by
\begin{equation}
  \label{eq:firstdiagr}
  \Gamma_- = \int_{V^{n+1}} d^{d(n+1)}\hat r_1 \rho(\hat r_1) 
  p( \vec r_1\,^0 - \vec r\,^0 ) \left[ 
  e^{-\sum_{a=1}^n V(\vec r_1\,^a - \vec r\,^a)} -1 \right]\ .
\end{equation}
If the above expression is plugged into the truncated Mayer-cluster
expansion, one obtains a 
non-linear integral equation for $P(\tau)$ (details are given in 
App. \ref{app:firstdiagr}) 
\begin{equation}
  \label{eq:saddle}
  1-q+q\int_0^\infty d\tau\ P(\tau) e^{-i\tau x}
  = \exp\left\{ - c q + c q \int_0^\infty d\tau\ P(\tau) 
  \exp\left(-i\frac{\kappa\tau}{2(\kappa+\tau)}x\right)  \right\}. 
\end{equation}
The above equation becomes exact, if we choose $p(\vec R_i^0 - \vec
R_j^0)=c/N$ to be distance-independent, instead of connecting only
particles, which are close in $d$-dimensional space. In fact, one can 
easily see that in the long-ranged case all higher diagrams in the 
Mayer-cluster expansion tend to zero in the thermodynamic limit, 
leaving only the first diagram.

By sending $x\to -i\infty$, we find a simple relation for the gel
fraction,
\begin{equation}
  \label{eq:gel}
  1-q = \exp\{- c q\}\ 
\end{equation}
first derived in the context of gelation in \cite{ZiGoGo}. It
coincides with the size of the giant component in random graph
percolation \cite{Erdoes}. Eq. (\ref{eq:gel}) always has the solution
$q=0$, corresponding to the fluid state with a vanishing fraction of
localized particles. At the critical value of the average
connectivity, $c_{crit}=1$, a second solution appears continuously,
accounting for the finite fraction of particles in the macroscopic
cluster.  If we slightly increase the number of crosslinks beyond the
percolation threshold, $c=1+\varepsilon$ ($0<\varepsilon\ll 1$), we
find $q = 2\varepsilon + O(\varepsilon^2)$, {\it i.e.} the gel
fraction grows linearly with the distance from the transition.

The equation for $P(\tau)$ can be simplified if we substitute
Eq. (\ref{eq:gel}) into Eq. (\ref{eq:saddle}), expand the right-hand
side, and invert the Fourier transform:
\begin{equation}
  \label{eq:iteration}
  P(\tau) = e^{-c q} \sum_{l=1}^\infty \frac{c^l q^{l-1}}{l!} 
  \int_0^\infty d\tau_1\cdots d\tau_l\ P(\tau_1)\cdots P(\tau_l)\ 
  \delta\left( \tau - \sum_{i=1}^l \frac{\kappa\tau_i}{\kappa+\tau_i} 
  \right)
\end{equation}
In the critical region $q$ is small so that the expansion on the right
hand side can be truncated.  The typical inverse squared localization
length grows linearly with $\varepsilon$. This can be seen by
rescaling it as $\tau=\varepsilon\kappa \theta$. We introduce the
scaling function $\pi(\theta)=\varepsilon\kappa
P(\varepsilon\kappa\theta)$ and expand (\ref{eq:iteration}) to second
order in $\varepsilon$ (see App. \ref{app:scalingfunction}) 
\begin{equation}
  \label{eq:scalingfunction}
  (1-2\theta) \pi(\theta) = \theta^2 \pi(\theta) 
  + \int_0^\theta d\theta_1\ \pi(\theta_1) \pi(\theta-\theta_1)\ . 
\end{equation}
The above Eqs. (\ref{eq:gel}) and (\ref{eq:scalingfunction}) were 
derived previously as the saddle-point
approximation for polymers crosslinked according to the Deam-Edwards
distribution \cite{Castillo}. In that work, a melt of linear
macromolecules was studied in contrast to the present work, which
starts from a fluid of point particles. Since the gelation transition
is a continuous phase transition with a diverging correlation length,
the critical behavior of the two systems is the same. This kind of
universality with respect to the building blocks has been noticed
previously \cite{Huthmann} and in fact also holds for different
crosslinking procedures.

The numerical solution of Eq.\ (\ref{eq:scalingfunction}) in the
critical region was given in Ref.\ \cite{Castillo}, here we go on to a
numerical solution of the full Eq.\ (\ref{eq:iteration}) for
arbitrary connectivities.  This numerical solution is based on a
replica-symmetric variant of the population dynamics introduced in
\cite{MePa2}. It starts with a large initial population ${\cal T}=\{
\tau_1,...,\tau_{\cal M}\}$ drawn randomly from some initial
distribution $P_0(\tau)$, and the gel fraction $q$ is determined from
Eq.\ (\ref{eq:gel}). The population is iterated in
the following way:\\
(i) Draw a random positive integer $l$ with probability
$e^{-c q} c^lq^{l-1}/l!$.\\
(ii) Choose $l+1$ random integers $i_0,i_1,...,i_l$ uniformly from
$\{1,...,{\cal  M}\}$.\\
(iii) Replace $\tau_{i_0}$ by $\sum_{k=1}^l
\kappa\tau_{i_k}/(\kappa+\tau_{i_k})$. \\
(iv) Go back to (i) with the updated population.\\
After a sufficiently large number of iterations, the histogram of
${\cal T}$ will be a good approximation of $P(\tau)$. For getting a
better statistics, a time average over many iterations steps can be
taken (starting after some waiting time needed for approaching a
stable fix point).

\begin{figure}[htb]
\begin{center}
\myscalebox{\includegraphics{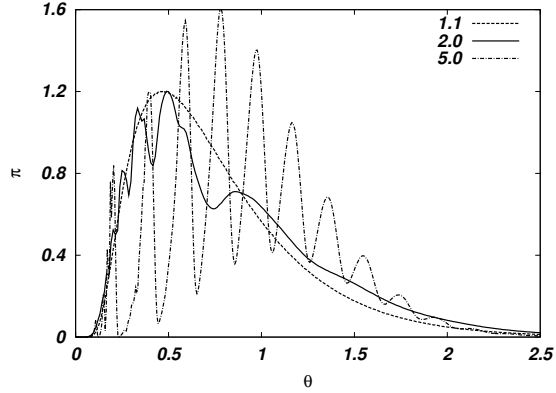}}
\end{center}
\caption{Rescaled distribution 
  $\pi(\theta)=(c-1)\kappa\,P((c-1)\kappa\theta)$ 
  of inverse squared localization lengths, for average
  connectivities $c = 1.1,\ 2.0,\ 5.0$, and the simplest truncation of
  the Mayer-cluster expansion. The 
  results are obtained numerically by the population dynamics for 
  ${\cal M}=10^6$ and $10^4 {\cal M}$ iteration steps. We have tested
  that the iteration has reached a stationary point. The curves show
  the crossover from the smooth scaling function for small
  connectivities to the peaked structure for large connectivities.}
\label{fig:poftau}
\end{figure}

We have implemented this algorithm, and calculated $P(\tau)$ for
several values $c=\rho_0 W_-$ of the average connectivity. The results
are shown in Fig. \ref{fig:poftau}. For connectivities close to the
percolation threshold, $P(\tau)$ in fact follows the scaling function
described by (\ref{eq:scalingfunction}), and then starts to deviate
and to develop a peak structure. Still, the typical values for the
inverse squared localization lengths scale approximately like $c-1$.
The peak structure observed for connectivities far beyond the gelation
point becomes more and more pronounced, and allows for an analytical
solution of the leading terms in $P(\tau)$ in the limit of high, but
finite connectivity.

We realize that $q$ approaches 1 exponentially for increasing $c$ and
the typical values of $\tau$ grow proportionally to $c$. Consequently 
we have
\begin{equation}
  \label{eq:highc}
  \sum_{i=1}^l \frac{\kappa\tau_i}{\kappa+\tau_i} = l\kappa  
  + {\cal O}\left(\frac{l\kappa}{c}\right) \ .
\end{equation}
Neglecting all subdominant terms, we find
\begin{equation}
  \label{eq:limitp}
  P(\tau) \simeq \sum_{l=0}^\infty e^{-c}\frac{c^l}{l!} \delta (\tau
  -l\kappa)\ .
\end{equation}
Including also subdominant terms, the peaks are shifted by ${\cal
O}(\kappa)$, whereas the width becomes ${\cal O}(c^{-1/2}\kappa)$,
leading to the structure observed in the numerical solution for
$P(\tau)$.

This leads to a very simple and attractive picture for strongly
crosslinked systems: The localization length of a particle depends to
leading order only on the number of crosslinks attached to the
particle, or more precisely, its inverse square localization length
equals the coordination times the crosslink strength. The
distribution of localization lengths  can then be written as a
superposition of contributions of all coordinations $l$, each
one occurring according to the  Poisson distribution of
Eq.\ (\ref{eq:connect}), $w(l)=c^l e^{-c}/l!$.

An interesting consequence of the above interpretation arises in
the context of the connectivity correlations discussed in section
\ref{sec:conncorr}: The same correlations transmit to the
localization length. A weakly localized particle is, on average,
surrounded by other weakly localized particles, a strongly localized
one by other strongly localized particles. This kind of local
inhomogeneities is also observed numerically for structural glasses,
see \cite{VoKoBiZi}. Even though this picture is based on the
simplest truncation of the Mayer-cluster expansion, we expect it to
hold in the full theory (see the discussion in the next section).

\subsection{First correction to mean-field}
\label{sec:triangle}

How does the inclusion of higher-order diagrams change the picture drawn
above? We will give a partial answer by including also the triangular
diagram $\Gamma_\Delta$ into the truncated Mayer series. The 
single-molecule density is then given by
\begin{equation}
  \label{eq:mayer3}
  \rho(\hat r) =\exp\left\{-\mu+\Gamma_- +\Gamma_\Delta \right\}\ .
\end{equation}
The exponent on the rhs of Eq. (\ref{eq:saddle}) has thus to be 
completed by the term
\begin{eqnarray}
  \label{eq:trianglen0}
  \Gamma_\Delta 
  &\mapsto& \frac{\rho_0^2 W_\Delta}{2} \int_0^\infty d\tau_1\ d\tau_2 
  \tilde P(\tau_1)\ \tilde P(\tau_2) \non\\
  && \times \sum_{\alpha_{1,2,3}\in\{0,\kappa\}}
  (-1)^{1+\frac{\alpha_1}\kappa+\frac{\alpha_2}\kappa+
  \frac{\alpha_3}\kappa} \exp\left\{ - i x
  \ \frac{\tau_1\tau_2(\alpha_1+\alpha_2)+(\tau_1+\tau_2)
    (\alpha_1\alpha_2+ \alpha_1\alpha_3+\alpha_2\alpha_3)}{
    (\tau_1+\alpha_1+\alpha_3)(\tau_2+\alpha_2+\alpha_3)
    -\alpha_3^2} \right\}
\end{eqnarray}
where we use the abbreviation 
$\tilde P(\tau) = (1-q) \delta(\tau) + q P(\tau)$
for the localization-length distribution of all $N$ particles,
including the delocalized ones with $\tau=0$. The constant
$W_\Delta$ is given in Eq. (\ref{eq:triangles}). Even if the
derivation is slightly more involved, it follows the ideas presented
in App. \ref{app:firstdiagr} for $\Gamma_-$. 

\begin{figure}[htb]
\begin{center}
\myscalebox{\includegraphics{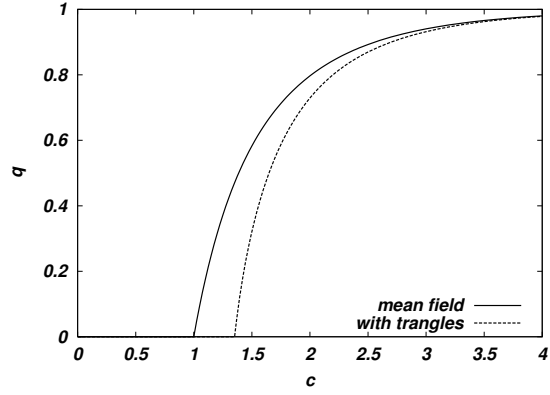}}
\end{center}
\caption{Gel fraction as a function of the average connectivity
  $c=\rho_0 W_-$ for $d=3$ and both truncations of the Mayer-cluster
  expansion. When the triangular diagram is included, the existence of
  a non-zero gel fraction sets in later.}
\label{fig:giant}
\end{figure}

In the limit $x\to -i\infty$, we obtain a closed equation for the
gel fraction $q$,
\begin{equation}
  \label{eq:gel3}
 1-q=\exp\{-c q +\rho_0^2W_\Delta q(1-q)\}\ . 
\end{equation}
Specializing to the Gaussian shape of $p(\vec r\,^0)$ given in
(\ref{eq:probability}), we find $\rho_0^2W_\Delta=3^{-d/2} c^2$. A
nonzero gel fraction appears continuously at the critical 
value $c_{crit}$ of the average connectivity given by 
\begin{equation}
  \label{eq:perc3}
 \gamma:=\frac{2}{ 3^{\frac{d}{2}}} c_{crit} =  1-
  \sqrt{1-4\cdot3^{-\frac{d}{2} }}  .
\end{equation}
For $d=3$ we find $c_{crit}=1.3515$ , and for increasing spatial
dimension the critical connectivity decreases exponentially to
one. So, in contrast to the simplest truncation, we find a
dimension-dependent gelation point. The increasing number of needed
crosslinks is due to the fact that the finite-dimensional disorder
produces more compact structures (triangles, tetrahedra, etc.), and
more links are needed to build up a macroscopic cluster of connected
particles, see also Fig. \ref{fig:giant}.

Increasing the connectivity $c$ by a small amount
$\varepsilon$ above the percolation threshold, $c=c_c+\varepsilon$, 
we can expand Eq. (\ref{eq:gel3}) in both $\varepsilon$ and
$q$. Neglecting contributions of ${\cal O}(\varepsilon^k q^l)$ with 
$k+l\geq 3$, we find
\begin{equation}
  \label{eq:criticalexp3}
  q = 2 \frac{1-\gamma}{1-\gamma c_{crit}} \varepsilon + 
  {\cal O}(\varepsilon^2)\ .
\end{equation}
The gel fraction $q$ starts to grow linearly with the distance
from the transition; the critical exponent for the gel fraction is
thus given by its mean-field value. Only the prefactor is changed, in
3 dimensions it equals 3.231. The prefactor approaches its mean-field 
value exponentially, as the spatial dimension is increased.

We expect the same critical exponent to be valid for any finite
truncation of the Mayer-cluster expansion, because the equation
determining the gel fraction has the form 
\begin{equation}
\label{eq:general}
1-q=\exp\left\{ \sum_{i=1}^{i_{\mathit max}} a_i q^i \right\}
\end{equation}
where $i_{max}$ is the maximal number of black vertices in the
considered diagrams. The gelation transition happens at $a_1=1$,
implying a linear growth of the gel fraction. A different exponent can
only result either from a resummation of an infinite series or if
accidentally $a_2=1/2$ at the critical point, a situation which
obviously would correspond to a very specific choice of the model
parameters.  

For general crosslink concentration $c$ and the truncation including
only the linear and triangular diagrams, the equation for the
distribution of localization lengths follows from Eqs.\
(\ref{eq:mayer3},\ref{eq:trianglen0}). It can be written as a series
expansion, similar to the mean-field result of Eq.\
(\ref{eq:iteration}). However, not all coefficients of the expansion
are positive. Hence, these coefficients cannot be interpreted as
probabilities any more, and the population dynamics cannot be applied.

We can still compute the distribution of localization lengths
in the critical regime. The scaling function has to be modified
according to
\begin{equation}
  \label{eq:rescale3}
  \pi(\theta) =
  \frac{1-\gamma}{1-\frac{2}{3}\gamma c_{crit}}\varepsilon\kappa \ 
  P\left(\frac{1-\gamma}{1-\frac{2}{3}\gamma
      c_{crit}}\varepsilon\kappa \theta \right)\ .
\end{equation}
Proceeding analogously to App. \ref{app:scalingfunction}, we reproduce
Eq.  (\ref{eq:scalingfunction}) and hence find that the critical
behavior of the distribution of inverse squared localization lengths
remains unchanged by the triangular diagram.

Also the limiting case of high crosslink densities $c\gg1$ is
tractable.  In particular, ansatz (\ref{eq:limitp}) still solves the
integral equation up to corrections of ${\cal O}(c^0)$. To this end,
we plug this ansatz into $\Gamma_\Delta$ as given in
(\ref{eq:trianglen0}), and use that $\tau_{1,2}={\cal O}(c)$ with a
probability that approaches one for increasing $c$.  Hence the
argument of the exponential in (\ref{eq:trianglen0}) is given by
$-i(\alpha_1+\alpha_2)x+{\cal O}(1/c)$ and does not depend on
$\alpha_3$ to leading order in $1/c$.  The summation over
$\alpha_3$ thus leads to a cancellation of the leading order, such
that $\Gamma_\Delta={\cal O}(1/c)$, {\it i.e.} the triangular diagram
does not contribute to the leading order in $1/c$ but only to the
corrections of solution (\ref{eq:limitp}). We expect the same to
hold also for higher-order diagrams.

More generally, we expect mean-field theory to give qualitatively
correct results in the gel phase away from the critical point. The
quantitative agreement becomes increasingly better for growing
connectivity, as can be seen for the gel fraction (see
Fig.\ref{fig:giant}) as well as for the distribution of localization
length.

\section{The Deam-Edwards distribution}
\label{sec:deamedwards}

Previous work on the statistical mechanics of gelation does not use
the cluster statistics of percolation theory, but instead follows the
elegant strategy of Deam and Edwards \cite{DeEd}, who have given an
implicit formula for the distribution of crosslinks, which is believed
to be equivalent to finite dimensional percolation. It is the aim of
this section to clarify the relation between the two approaches and
put the Deam-Edwards distribution in a more general context.

We start from Eq. (\ref{eq:disorder2}), but allow for fluctuations in 
the total number of crosslinks, assuming a Poisson distribution.  A 
particular crosslink configuration is then constructed in three steps:

1) Choose a configuration $R^0$ randomly, according to the distribution 
\begin{equation}
\label{eq:distr1}
\varphi(R^0) =\frac{e^{-\psi(R^0)}}{\int d^{dN} R\ e^{-\psi(R)} }. 
\end{equation}
A natural choice would be $\psi(R^0)=U(R^0)$ in order to model
instantaneous crosslinking of a fluid with correlations induced by
the pair interactions $U(R^0)$. As we shall see, the Deam-Edwards
distribution does not correspond to this choice.

2) Given the configuration $R^0$, choose the total number of
crosslinks $M$ according to a Poisson distribution
\begin{equation}
W(M|R^0)=\frac{h(R^0)^M}{M!}e^{-h(R^0)}\ ,
\end{equation}
where $h(R^0)$ is an arbitrary positive function.

3) Given the configuration $R^0$ and the total number of
crosslinks $M$, choose a crosslink configuration ${\cal
  C}=\{(i_e,j_e)\}_{e=1}^M$   with probability
\begin{equation}
\label{eq:distr2}
P({\cal C}|R^0,M)=\prod_{e=1}^{M} p(i_e,j_e|R^0,M)\ .
\end{equation}

 An observable $f(R^0,{\cal C})$, which depends not
only on the crosslink configuration ${\cal C}$  but possibly also 
on $R^0$ is then averaged according to
\begin{equation}
\label{eq:genaverage}
\overline{f(R^0,{\cal C})}=
\sum_{M=1}^{\infty}\,\prod_{e=1}^M\sum_{i_e,j_e=1}^N \,\int d^{dN}R^0
\varphi(R^0) \frac{(h(R^0))^M}{M!} e^{-h(R^0)}
P({\cal C}|R^0,M) f(R^0,{\cal C})\ .
\end{equation}

Modeling continuum percolation, we take $\varphi(R^0)=1/V$ and 
$h(R^0)=c N /2$ independent of
$R^0$, so that the mean number of crosslinks $\overline{M}=c N/2$ is
macroscopic and fluctuations around the average are small. The number
of crosslinks per particle fluctuates around its finite mean $c$. 

In the approach of Deam and Edwards, one argues that particles which
have a high probability to be close in the fluid phase also have a
high probability to be crosslinked. The corresponding distribution of
crosslinks is defined implicitly through the average
of an observable \cite{DeEd,BaEd}, according to
\begin{equation}
\label{eq:deamedwards}
\overline{f(R^0,{\cal C})}^{DE}=\sum_{M=1}^{\infty} \frac{(\mu)^M}{M!}  
\prod_{e=1}^M\sum_{i_e,j_e=1}^N \frac{\int d^{dN} R^0 e^{-U(R^0)} 
\prod_{e=1}^M \Delta(|\vec R_{i_e}^0-\vec R_{j_e}^0|)f(R^0,{\cal C})}
{\int d^{dN} R^0 e^{-U(R^0)+\mu\sum_{i,j=1}^N \Delta(|\vec
    R_{i}^0-\vec R_{j}^0|)}} 
\end{equation}
Here $U(R^0)$ denotes the pair potential, acting among the particles of
the fluid, and $\Delta(x)$ is a short-ranged, positive function.  
The Deam-Edwards distribution is a special case of the above more
generally defined distribution of crosslinks
(\ref{eq:distr1}-\ref{eq:distr2}), implying a special choice for 
$\varphi(R)$ and $h(R)$ denoted by $\varphi^{DE}(R)$ and $h^{DE}(R)$.
To uniquely identify these functions
we choose $f(R^0,{\cal C})=\delta(R^0-R)\delta(M-K)$ 
and compute its average according to Eq.(\ref{eq:deamedwards})
\begin{equation}
\overline{\delta(R^0-R)\delta(M-K)}^{DE}=\frac
{e^{-U(R)}}{\int d^{dN}R^0\quad e^{-U(R^0)
+\mu\sum_{i,j=1}^N \Delta(|\vec R_{i}^0-\vec R_{j}^0|)} }\
\frac{\left[ \mu \sum_{i,j=1}^N \Delta(|\vec R_{i}-\vec
    R_{j}|)\right]^K}{K!} \ .
\end{equation}
Computing the same average with the general class of distributions
according to Eq.\ (\ref{eq:genaverage})
\begin{equation}
\overline{\delta(R^0-R)\delta(M-K)}=\frac
{e^{-\psi(R)}} {\int d^{dN}R^0\quad e^{-\psi(R^0)} }
\frac{h(R)^K}{K!}
e^{-h(R)}
\end{equation}
allows us to identify
\begin{equation}
\psi^{DE}(R)=U(R)-\mu\sum_{i,j=1}^N \Delta(|\vec R_{i}-\vec
  R_{j}|) \qquad \mbox{and} \qquad
h^{DE}(R)=\mu\sum_{i,j=1}^N \Delta(|\vec R_{i}-\vec R_{j}|).
\end{equation}

We note that, first, the chemical potential of the crosslinks, which
determines the total number of crosslinks, depends on the disorder
configuration.  A short ranged $\Delta(|\vec R|)$ will give rise to an
intensive concentration, because $\sum_{i,j=1}^N\Delta(|\vec
R_{i}-\vec R_{j}|)$ should be of order ${\cal O}(N)$. Fluctuations are
expected to be of order ${\cal O}(\sqrt{N})$, so that one might hope
that in the macroscopic limit it does not matter, whether $h(R)$ in
the Poissonian distribution is taken to be constant or chosen to be
$R$-dependent as by Deam and Edwards.

Second, the potential $\psi^{DE}(R)$, which determines the crosslink
configuration, is {\it not} the potential of the underlying fluid as
it should be for chemical gelation. In particular the correlations of
the fluid should not depend on the function $\Delta(R)$ which
determines the probability of a crosslink to be formed. The choice of
Deam and Edwards has, however, technical advantages, because it
simplifies replica calculations: The replica theory resulting from the
above average (\ref{eq:deamedwards}) with $f({\cal C})=-\ln Z({\cal
C})$ is the only one which is symmetric with respect to permutations
of all $n+1$ replicas. In view of the above identifications, this
seems to be an artifact of the Deam-Edwards distribution and the
generic theory of gelation will not have this additional symmetry, but
will only be symmetric with respect to permutations of the $n$ thermal
replicas, as discussed in Sec.\ \ref{sec:op}.

\section{Conclusion}
\label{sec:conclusion}

We have studied the gelation transition as well as the highly
connected gel phase for crosslink distributions of $d$-dimensional
percolation. The average over the random connectivity can be achieved
with help of the replica trick. The resulting $n$-fold replicated,
effectively uniform theory is interpreted as a molecular fluid, such
that each particle of the unaveraged system corresponds to an
effective molecule with $(n+1)$ constituents. The uniform theory is
symmetric with respect to permutations of $n$ replicas, which are
introduced to represent $\ln Z$ and are called thermal replicas. One
additional replica is used to generate short-range connectivity
correlations and, in general, cannot be permuted with any of the
thermal replicas.

The molecule density entails all information about the order parameter
of the gel phase. "Bound" molecules indicate localization of the
particles, in the sense that copies (or thermal replicas) of the
original system are close to each other in real space. More precisely,
the order parameter of the gel phase is the
distribution of local static density fluctuations. This distribution
is non-Gaussian due to the finite connectivity per particle. Hence
moments of arbitrary order $l$ are necessary to specify the state of
the gel. Integrating the molecule density over all but $l$ of the $n$
thermal replicas, yields the $l^{th}$ moment of the local static
density fluctuations.

The effective molecular fluid allows for an analysis
in the framework of liquid-state theory. In particular a Mayer-cluster
expansion can be set up for the local molecule density. The
lowest-order term in the expansion yields back the results of
mean-field theory. Assuming a replica-symmetric solution, we can
compute corrections to mean-field theory. Here we do not concentrate
on non-classical, critical behavior, which is difficult to obtain in
such an expansion. Instead we focus on the properties of the gel phase
away from the critical point. Within mean-field theory the structure
of the gel is completely characterized by the distribution of
localization lengths $P(\xi^2)$, which has been computed previously
\cite{Castillo} in the critical region. Here we compute $P(\xi^2)$ for
all connectivities and point out a one-to-one correspondence between
the connectivity of a particle and its localization length. For high
connectivities the distribution of localization lengths is shown to
exhibit a multi-peak structure with the weight of each peak given by
the Poisson statistics of connectivities.

Corrections to mean-field theory increase the percolation threshold,
but do not change the qualitative picture of mean-field theory. In the
limit of increasing connectivity the corrections to mean-field theory
become less and less significant.

The analysis may possibly be extended to study two-point-correlation
functions or even higher correlations. Given the connectivity
correlations of neighboring sites, as discussed in
Sec.\ \ref{sec:conncorr}, we expect to find similar correlations for
strongly, respectively weakly localized particles. Another possible
extension of our work concerns the elastic properties of the gel. It
would certainly be interesting to compute the elastic constants within
a Mayer-cluster expansion.

Besides the Mayer-cluster expansion, we have discussed a rather
general class of crosslink distributions. This general framework helps 
to put the Deam-Edwards distribution in the context of
percolation theory, and allows to study a variety of connectivity
distributions. Whether or not the critical behavior is universal with
respect to the distribution of the disorder is unknown and so far has
hardly been addressed systematically in disordered system.

Furthermore, the Deam-Edwards distribution was shown to be the only
crosslink distribution giving rise to an average free energy which is
symmetric with respect to permutations of $n+1$ replicas. The generic
case - including percolation statistics as well as crosslinks with
correlations of the melt - is symmetric with respect to permutations
of the $n$ thermal replicas only. In view of these results, it might
be interesting to reconsider the issue of replica-symmetry breaking,
which was shown to be absent in gels with the Deam-Edwards
distribution \cite{GoZi2}.

\section*{Acknowledgment} We are grateful to Peter M\"uller for
numerous comments on the manuscript. The work of AZ has been supported
by the DFG through Grant No. Zi209/6-1 and SFB 1871.

\appendix

\section{Equation for $P(\tau)$ for the simplest truncated series}
\label{app:firstdiagr}

In this appendix, we derive the self-consistent equation for $P(\tau)$
from the simplest non-trivial truncation of the Mayer-cluster
expansion for the $l^{th}$ moment
\begin{equation}
  \label{A:saddle1}
  \rho^{(l)}(\vec r\,^1,...,\vec r\,^l) = \lim_{n\to 0} \int_V d^d
  r^0 \prod_{a=l+1}^n \int_V d^d r^a
  \exp \left\{ -\mu + \Gamma_-  \right\}\ .
\end{equation}
We compute $\Gamma_-$ with the ansatz of Eq.\ (\ref{eq:rs})
\begin{equation}
  \label{A:rs}
  \rho(\hat r) = (1-q)\frac{\rho_0}{V^n} +
  q \rho_0  \int d^d R \int_0^\infty d\tau_0 \ d \tau\ 
  P(\tau_0, \tau)\ \left( \frac{\tau_0}{2\pi} \right) 
  ^\frac{d}{2} \left( \frac{\tau}{2\pi} 
  \right)^\frac{dn}{2} \exp\left\{ - \sum_{a=0}^n\frac{\tau_a}{2}
  (\vec r\,^a-\vec R)^2\ ,
  \right\}
\end{equation}
where we have introduced $\tau_a := \tau,$ for all $a=1,...,n$.
The ansatz is first simplified by integrating over $\vec R$
\begin{equation}
  \label{A:rs1}
  \rho(\hat r) = (1-q)\frac{\rho_0}{V^n} +
  q \rho_0 \int_0^\infty d\tau_0 \ d \tau\ 
  P(\tau_0, \tau)\ \left( \frac{\tau_0}{\tau_0+n\tau} \right) 
  ^\frac{d}{2} \left( \frac{\tau}{2\pi} 
  \right)^\frac{dn}{2} \exp\left\{ -\frac{1}{4(\tau_0+n\tau)}
  \sum_{a,b=0}^n \tau_a \tau_b (\vec r\,^a-\vec r\,^b)^2
  \right\}\ ,
\end{equation}
and then it is plugged into the expression for $\Gamma_-$
\begin{equation}
  \label{A:def}
  \Gamma_- = \int d^{d(n+1)}\hat r_1 \rho(\hat r_1) 
  p( \vec r_1\,^0 - \vec r\,^0 ) \left[ 
  e^{-\sum_{a=1}^n V(\vec r_1\,^a - \vec r\,^a)} -1 \right]\ .
\end{equation}
We use harmonic interactions $V(\vec r)$ and a Gaussian-shaped
$p( \vec r)$ to find 
\begin{eqnarray}
  \label{A:calc1}
  \Gamma_- &=& - \rho_0 W_- + \rho_0 (1-q) \int \frac{d^{d(n+1)}\hat
    r_1}{V^n} \  p( \vec r_1\,^0 - \vec r\,^0 ) 
  \exp\left\{-\frac\kappa 2 \sum_a (\vec r_1\,^a - \vec r\,^a)\right\} 
  \non\\
  && + \rho_0 q \int d^d R \int_0^\infty d\tau_0 \ 
    d \tau\ P(\tau_0, \tau) 
    \left( \frac{\tau_0}{2\pi} \right)^\frac{d}{2} 
    \left( \frac{\tau}{2\pi}\right)^\frac{dn}{2} 
    \int d^{d(n+1)} \hat r_1\  \non\\
  &&\times\exp\left\{ 
   -\frac{\tau_0}{2}(\vec r_1\,^0 - \vec R)^2 - \frac{\tau}{2} 
   \sum_{a=1}^n (\vec r_1\,^a-\vec R)^2 -\frac{a}2 (\vec r_1\,^0
   - \vec r\,^0)^2 -\frac \kappa 2 \sum_{a=1}^n (\vec r_1\,^a 
   - \vec r\,^a)^2 \right\}\ .
\end{eqnarray}
The integrals over $\hat r_1$ are Gaussian and hence can be performed. 
We are interested in the leading term in the limit $n\to 0$, and hence
 replace $C^n\to 1$ with $C$ an arbitrary constant. In this way we obtain
\begin{eqnarray}
  \label{A:calc3}
  \Gamma_- &\to& - \rho_0 W_- q + \rho_0 q 
    \int d^d\vec R \int_0^\infty d\tau_0 \ d\tau\ P(\tau_0, \tau) 
    \left( \frac{\tau_0}{a+\tau_0} \right)^\frac{d}{2}\exp\left\{ 
   -\frac{a\tau_0}{2(a+\tau_0)} (\vec r\,^0 - \vec R)^2 -
   \frac{\kappa\tau}{2(\kappa+\tau)} \sum_{a=1}^n 
  (\vec r\,^a-\vec R)^2 \right\} \non\\
  &\to& - c q + c q 
  \int_0^\infty d\tau_0 \ d\tau\ P(\tau_0, \tau) \left( 
  \frac{a\tau_0}{2\pi(a+\tau_0)} \right)^\frac{d}{2} \int_0^\infty 
  d\sigma_0\ \delta\left( \sigma_0-\frac{a\tau_0}{a+\tau_0} \right)
  \int d\sigma\ \delta\left( \sigma-\frac{\kappa\tau_0}{\kappa+\tau}
  \right) \non\\
  &&\times \int d^d R \quad
   \exp\left\{ -\frac{\sigma_0}2 (\vec r\,^0 - \vec R)^2 -
   \frac\sigma 2 \sum_{a=1}^n  (\vec r\,^a-\vec R)^2 \right\}\ .\non
\end{eqnarray}
The $\vec R$-integration can be carried out, yielding
\begin{eqnarray}
  \label{A:calc4}
  \Gamma_- &\to& - c q + c q \int_0^\infty d\tau_0 \ d\tau\ P(\tau_0,
  \tau) \left( \frac{a\tau_0}{2\pi(a+\tau_0)} \right)^\frac{d}{2}
  \int_0^\infty d\sigma_0\ \delta\left(
  \sigma_0-\frac{a\tau_0}{a+\tau_0} \right) \int_0^\infty d\sigma\
  \delta\left( \sigma-\frac{\kappa\tau_0}{\kappa+\tau} \right) \non\\
  &&\times \left( \frac{2\pi}{\sigma_0+n\sigma }\right)^{\frac d 2}
   \exp\left\{ -\frac{1}{4(\sigma_0+n\sigma)} \sum_{a,b=0}^n  
   \sigma_a \sigma_b (\vec r\,^a-\vec r\,^b)^2 \right\}\non\\
  &\to& - c q + c q 
  \int_0^\infty d\tau_0  d\tau\ P(\tau_0, \tau) \int 
  d\sigma_0\ \delta\left( \sigma_0-\frac{a\tau_0}{a+\tau_0} \right)
  \int d\sigma\ \delta\left( \sigma-\frac{\kappa\tau}{\kappa+\tau}
  \right) \exp\left\{ -\frac{1}{4 \sigma_0} \sum_{a,b=0}^n  
   \sigma_a \sigma_b (\vec r\,^a-\vec r\,^b)^2 \right\}\non\ .
\end{eqnarray}
We have introduced $\sigma_a=\sigma$ for $a=1,...,n$ and 
have removed trivial $n$-dependencies which are irrelevant in the
replica limit $n\to 0$. 

In order to obtain an equation for $P(\tau)$, the above expression is
plugged into Eq.\ (\ref{A:saddle1}). For the sake of clarity, we only
consider $l=2$ 
\begin{equation}
  \label{A:saddle2}
  \rho^{(2)}(\vec r\,^1,\vec r\,^2) = \lim_{n\to 0} \int_V d^d
  r^0 \prod_{a=3}^n \int_V d^d r^a \exp \left\{ -\mu + \Gamma_-  
  \right\}\ .
\end{equation}
and note that all higher moments lead to the same equation for
$P(\tau)$.  
The integrations over $\vec r\,^0$ and $\vec r\,^a,\ a=3,...,n,$ can
be carried out by expanding the exponential on the rhs of Eq.\  
(\ref{A:saddle2}). The result
\begin{eqnarray}
  \label{A:res}
  \rho^{(2)}(\vec r\,^1, \vec r\,^2) &=& e^{-\mu-cq} 
  \sum_{l=0}^\infty  \frac{ (cq)^l }{l!} 
  \int_0^\infty d\tau_1 \cdots d\tau_l P(\tau_1)\cdots
  P(\tau_l) \int_0^\infty d\sigma \delta\left(\sigma-  \sum_{i=1}^l 
  \frac{\kappa\tau_i}{\kappa+\tau_i} \right) \non\\
  &&\times \left(\frac\sigma{4\pi} \right)^{\frac d 2} \exp\left\{ 
  -\frac \sigma 4 (\vec r\,^1 -\vec r\,^2)^2  \right\} \non
\end{eqnarray}
depends only on the reduced distribution $P(\tau)=\int d\tau_0\ 
P(\tau_0,\tau)$ of inverse squared localization lengths of the 
crosslinked system. For the left hand side of the last equation we use
expression (\ref{eq:physloc}) and compare coefficients of
Gaussians of the same variance. In the last step, we adjust
$\mu$ to fix the normalization and find
\begin{equation}
  \label{A:res2}
  (1-q)\delta(\tau)+q P(\tau) =  e^{-cq} \sum_{l=0}^\infty
  \frac{ (cq)^l }{l!} \int_0^\infty d\tau_1 \cdots d\tau_l 
  P(\tau_1)\cdots P(\tau_l) \delta\left(\tau -  \sum_{i=1}^l 
  \frac{\kappa\tau_i}{\kappa+\tau_i} \right)\ .
\end{equation}
Taking the Fourier transform with respect to $\tau$, allows us to
evaluate the sum on the rhs of Eq.\ (\ref{A:res2}) and leads to Eq.\
(\ref{eq:saddle}).

\section{Scaling function for the truncated series}
\label{app:scalingfunction}

In order to derive the scaling function for the distribution of
localization lengths in the critical region, we start from the full
equation (\ref{eq:iteration}) for the simplest truncated Mayer 
expansion, or equivalently for the mean-field theory:
\begin{equation}
  \label{B:iteration}
  P(\tau) = e^{-c q} \sum_{l=1}^\infty \frac{c^l q^{l-1}}{l!} 
  \int_0^\infty d\tau_1\cdots d\tau_l\ P(\tau_1)\cdots P(\tau_l)\ 
  \delta\left( \tau - \sum_{i=1}^l \frac{\kappa\tau_i}
  {\kappa+\tau_i} \right)
\end{equation}
The critical connectivity is given by $c=1$. If we further increase 
the number of crosslinks, $c=1+\varepsilon$ ($0<\varepsilon\ll
1$), a macroscopic gel component appears. Its  fraction is given by
\begin{equation}
  \label{B:critical2}
  q = 2\varepsilon + O(\varepsilon^2)\ .
\end{equation}
Moreover, the inverse squared localization lengths, i.e. the $\tau$'s, 
are expected to grow linearly with the distance from the critical
point. We thus rescale $P(\tau)$ by setting 
$\tau = \varepsilon\kappa\theta$, and define the distribution
\begin{equation}
  \label{B:scaling}
  \pi(\theta) = \varepsilon\kappa P( \varepsilon\kappa \theta)\ .
\end{equation}
Plugging this into equation (\ref{B:iteration}), we find
\begin{eqnarray}
  \label{B:calc}
  \pi(\theta) &=& e^{-c q} \sum_{l=1}^\infty \frac{c^l q^{l-1}}{l!} 
  \int d\theta_1\cdots d\theta_l\ \pi(\theta_1)\cdots \pi(\theta_l)\ 
  \delta\left( \theta - \sum_{i=1}^l \frac{\theta_i}
  {1+\varepsilon \theta_i} \right) \non\\
&=& (1-2\varepsilon)(1+\varepsilon) \int
  d\theta_1\ \pi(\theta_1)\ \delta\left(\theta -
  \theta_1(1-\varepsilon \theta_1)+O(\varepsilon^2)\right) \non\\
&&+ (1-2\varepsilon)(1+\varepsilon)^2\varepsilon \int
  d\theta_1 d\theta_2\ \pi(\theta_1)\  \pi(\theta_2)\ \delta\left( 
  \theta - \theta_1(1-\varepsilon \theta_1)-
  \theta_2(1-\varepsilon \theta_2) +O(\varepsilon^2) \right)
  +O(\varepsilon^2)\non\\
&=& (1-\varepsilon)  \int d\theta_1\ \pi(\theta_1)\ \delta\left( 
  \theta - \theta_1(1-\varepsilon \theta_1) \right) 
  + \varepsilon  \int d\theta_1 d\theta_2\ \pi(\theta_1)\  
  \pi(\theta_2)\ \delta( \theta - \theta_1- \theta_2)
  +O(\varepsilon^2)\non\\ 
&=& \pi(\theta) + \varepsilon\left\{-\pi(\theta) +2\theta \pi(\theta) 
  +\theta^2\pi'(\theta)
  + \int d\theta_1\ \pi(\theta-\theta_1)\pi(\theta)  \right\}
  +O(\varepsilon^2)\ .\non  
\end{eqnarray}
The first non-trivial order thus gives the expected
integro-differential equation (\ref{eq:scalingfunction}) for the
scaling function  $\pi(\theta)$.



\begin{thebibliography}{99}
\bibitem{FlSt} Classical contributions are described in P.J. Flory,
  {\it Principles of Polymer Chemistry}, (Cornell University Press,
  Ithaca, 1953).
\bibitem{St} D. Stauffer, J. Chem. Soc. Faraday Trans. II {\bf 72},
  1354 (1976).
\bibitem{Ge} P.G. de Gennes, {\it Scaling Concepts in Polymer
    Physics}, (Cornell University Press, Ithaka, 1979).
\bibitem{Adam} M. Adam, M. Delsanti, J.P. Munch, and D. Durand,
  J. Phys. (Paris) {\bf 48}, 1809 (1987).
\bibitem{Ge2} P.G. de Gennes, J. Phys. (Paris) {\bf 38}, L-355
  (1977). 
\bibitem{DeEd} R.T. Deam, S.F. Edwards, Phil. Trans. R. Soc. A {\bf
    280}, 317 (1976).
\bibitem{BaEd} R.C. Ball and S.F. Edwards, Macromolecules {\bf 13},
  748 (1980).
\bibitem{GoGo} P.M. Goldbart and N. Goldenfeld, Phys. Rev. Lett. {\bf
    58}, 2676 (1987); Phys. Rev. A {\bf 39}, 1402 (1989); {\it ibid.}
  {\bf 39}, 1412 (1989).  
\bibitem{Pa} S.V. Panyukov, JETP Lett. {\bf 55}, 608 (1992).
\bibitem{PaRa} S.V. Panyukov and Y. Rabin, Phys. Rep. {\bf 269}, 1 
    (1996). 
\bibitem{GoZi} P.M. Goldbart and A. Zippelius, Phys. Rev. Lett. {\bf
    71}, 2256 (1993).
\bibitem{Castillo} H. E. Castillo, P. M. Goldbart, and A. Zippelius,
  Europhys. Lett. {\bf 28}, 519 (1994).
\bibitem{CaGoZi} P.M. Goldbart, H. Castillo, and A. Zippelius,
  Adv. Phys. {\bf 45}, 393 (1996).
\bibitem{Kane} W. Peng and P. M. Goldbart, Phys. Rev. E
  {\bf 61}, 3339 (2000); W. Peng, P. M. Goldbart, and A. J. McKane, 
  Phys. Rev. E {\bf 64}, 031105 (2001).
\bibitem{Jansen} H. K. Janssen and O. Stenull, Phys. Rev. E {\bf 64},
  026119 (2001).
\bibitem{Vilgis} M. P. Solf and T. A. Vilgis, J. Phys. A {\bf 28},
  6655 (1995). 
\bibitem{Bo} B. Bollobas, {\it Random graphs}, (Academic Press,
  London, 1985).
\bibitem{Erdoes} {P. Erd{\H o}s and A. R\'enyi},
  {Magyar Tud. Akad. Mat. Kut. Int. K{\H o}zl.}  {\bf 5}, {17} (1960); 
  reprinted in: {The Art of Counting}
  edited by {J. Spencer} (MIT Press, Cambridge, MA, 1973)
\bibitem{Edwards} S. F. Edwards, in {\it Polymer networks}, ed. by
  A. J. Chrompff and S. Newman (Plenum, New York, 1971).
\bibitem{Go} P.M. Goldbart, J. Phys. Cond. Matt. {\bf 12}, 6585
  (2000). 
\bibitem{Cr} C.A. Croxton, {\it Liquid state physics - A statistical
    mechanics introduction}, (Cambridge University Press, 1974).
\bibitem{MoHi} T. Morita and K. Hiroike, Prog. Theor. Phys. {\bf 25},
  537 (1961).
\bibitem{ZiGoGo} A. Zippelius, P. M. Goldbart, and N. Goldenfeld,
  Europhys. Lett. {\bf 23}, 451 (1993).
\bibitem{Huthmann} M. Huthmann, M. Rehkopf, A. Zippelius and
  P. Goldbart, Phys. Rev. E {\bf 54}, 3943 (1996). 
\bibitem{MePa2} M. M\'ezard and G. Parisi, Eur. Phys. J. B {\bf 20}, 
  217 (2001).
\bibitem{VoKoBiZi} K. Vollmayr-Lee, W. Kob, K. Binder, and
  A. Zippelius, preprint cond-mat/0109460.
\bibitem{GoZi2} P.M. Goldbart and A. Zippelius, J. Phys. A {\bf 27},
  6375 (1994).
\end{thebibliography}
\end{document}